\documentclass[letterpaper, 10 pt, conference]{ieeeconf}  % Comment this line out
\IEEEoverridecommandlockouts                              % This command is only
\usepackage{authblk}                                                       % needed if you want to
\usepackage{graphicx}                                                        % use the \thanks command
\title{\LARGE \bf
Insights from Statistical Analysis of Opioid Data
}

\author[1]{Kaustav Basu}
\author[1]{Sandipan Choudhuri}
\author[1]{Arunabha Sen}
\author[2]{Aniket Majumdar}
\affil[1]{NetXT Lab, School of Computing, Informatics and Decision Systems Engineering, Arizona State University}
\affil[2]{Galvanize, Inc.}
\affil[1]{\{kaustav.basu, schoud13, asen\}@asu.edu}
\affil[2]{aniket.majumdar@galvanize.com}

\begin{document}

\maketitle
\thispagestyle{empty}
\pagestyle{empty}

%%%%%%%%%%%%%%%%%%%%%%%%%%%%%%%%%%%%%%%%%%%%%%%%%%%%%%%%%%%%%%%%%%%%%%%%%%%%%%%%
\begin{abstract}
Opioid overdose has emerged as a full blown epidemic in the United States. 
In the last few years, there has been an alarming increase in Opioid related deaths, resulting in the loss of 63,600 lives in 2016 alone. The epidemic which is killing more than 100 people each day, was declared as a public health emergency by the US government, in October 2017. Although a few health related companies and commercial firms have examined this important issue from various available data sources, to the best of our knowledge,  the academic community has not been engaged in research in this important topic. It can be safely noted that the study of the epidemic, from the data analytics perspective, is in its infancy. Given that a significant amount of Opioid related data is available in public domain, it provides the academic community an opportunity to analyze such data to provide recommendations to the public health authorities to mitigate the impact of the epidemic. In that vein, we collected some publicly available data to analyze the important contributing factors of the epidemic. In particular, we examine the role of the individuals prescribing Opioid drugs on the spread of the epidemic. In addition, we examine the impact of income level, age and educational level of various neighborhoods in a large US city, on Opioid related incidences, to find any correlation between them.

\end{abstract}

%%%%%%%%%%%%%%%%%%%%%%%%%%%%%%%%%%%%%%%%%%%%%%%%%%%%%%%%%%%%%%%%%%%%%%%%%%%%%%%%
\section{Introduction}
\label{intro}
Opioids are drugs, prescribed by health professionals to relieve patients from pain. Unfortunately, these drugs often lead to addiction. This addiction has emerged as a full blown epidemic in the United States. 
In the last few years, there has been an alarming increase in Opioid related deaths, resulting in the loss of 63,600 lives in 2016 alone. In October 2017, the epidemic was declared as a public health emergency by the US government \cite{emergency}. Although a few health related companies and commercial firms have examined this important issue from various available data sources, to the best of our knowledge,  the academic community has not been engaged in research in this important topic. Arguably, the study of the epidemic from the data analytics perspective, is in its infancy. Given that a significant amount of Opioid related data is available in public domain, it provides the academic community an opportunity to analyze such data to provide recommendations to the public health authorities to mitigate the impact of the epidemic. In that vein, we collected some publicly available data to analyze the important contributing factors of the epidemic. In particular, we examine the role of the individuals prescribing Opioid drugs on the spread of the epidemic. In addition, we examine the impact of income level, age and educational level of various neighborhoods in a large US city, on Opioid related incidences, to find any correlation between them. 

In the last few years, a small number of health and commercial companies have undertaken studies on Opioid related incidences, involving data analytic techniques. Blue Cross Blue Shield \cite{bcbs} for one, stated in their 2017 report that 21$\%$ of their commercially insured members filled at least one opioid prescription in 2015. Their data shows that members, with an opioid use disorder diagnosis, grew to $493\%$ over a seven year period, from 2010 to 2016. Their report also summarizes that women, over 45, have higher Opioid overdose rates than their counterparts in the same age bracket. On the other hand, men, under 45 have higher overdose rates than women under 45. Finally, they report that the Opioid overdose treatment rates are lower in the Southern states and in parts of the Midwest.

The Centers for Medicare and Medicaid Services \cite{cms}, an agency of the U.S. Department of Health and Human Services (HHS), maintains almost 24 million Opioid related prescriptions, written by 1 million unique health professionals (prescribers), in the U.S in 2014. The details of the data provided in these prescriptions are described in Section. \ref{data}. A small subset of this dataset with 25,000 unique prescribers, is available on  \cite{kaggle}. 

Data Science researchers from IBM Research and experts IBM Watson Health have recently embarked on applying data analytics and machine learning techniques to uncover new insights to address the opioids problem \cite{IBM}. Their effort is directed towards the analysis of the relationship between factors surrounding an initial opioid prescription, and a subsequent diagnosis of addiction. The goal of this research is to identify causal factors that lead to addiction diagnosis, taking into account all the variables associated with the initial prescription, such as opioid class, quantity, and related medical procedures and diagnoses. Some other efforts in this direction include, Mackey's study \cite{mackey} on illegal sales of prescription opioids online, utilizing Twitter data. Chary et. al. in \cite{chary} also analyzed Twitter data with a goal of identifying the location of the Opioid related Tweet.

In this paper, we try to analyze the important contributing factors of the epidemic from publicly available datasets. In particular, we attempt to provide answers to the following questions, 
\begin{itemize}
\item Q1: Is there a correlation between the prescribers, prescriptions and opioid related deaths in U.S. states?
\item Q2: Which prescribers are likely to prescribe more than 10 Opioid related prescriptions in a year?
\item Q3: Is there a correlation between the income level and Opioid related incidences, in a neighborhood?
\item Q4: Is there a correlation between the age and Opioid related incidences, in a neighborhood?
\item Q5: Is there a correlation between the education level and Opioid related incidences, in a neighborhood?
\end{itemize}

Our analysis shows a moderate level of correlation between Opioid related incidences with both the number of prescribers and the number of  prescriptions. Researchers from IBM \cite{IBM} also examined the question of ``Which prescribers are likely to prescribe more than 10 Opioid related prescriptions in a year?'' Using multiple machine learning algorithms, they computed accuracy of their predictions with values ranging from 60$\%$ to 84$\%$ \cite{IBMGit}. Treating the IBM accuracy results as the benchmark accuracy, we used boosting algorithms, to reach an accuracy of 85$\%$, a multilayer perceptron, to get an accuracy of 89$\%$, and a random forest classifier, which also had an accuracy of $89\%$. Our perceptron model did not take into account the specialty of the prescribers, but produced a model with better decision boundary. After analyzing the role of prescribers in the country as a whole, we examine the role of prescribers by state. We see a higher Opioid prescription rate in the southern states. We dive deeper and analyze the Opioid Prescription rate by prescriber specialty. We illustrate the top 10 Opioid prescribing specialties, in Fig. \ref{meds}. Finally, our analysis found a small negative correlation between Opioid related deaths with income, age, education level, when considered separately. Our analysis is presented in Section \ref{approach}.

\section{Datasets for Analysis}
\label{data}
In order to answer to the questions listed in the previous section, we first collected data from multiple sources and then munged the collected data to create additional datasets. The details of our data collection and data munging are provided in Sections \ref{collection} and \ref{munging}.

\subsection{Data Collection}
\label{collection}
Our collected data comprises of four different datasets $DS_1$, $DS_2$, $DS_3$ and $DS_4$. In the following we describe each one of them.\\ 

\textbf{$DS_1$}: It is the U.S. Opiate Prescriptions/Overdoses dataset available on \cite{kaggle}. This dataset comprises of 25000 unique prescribers, across the U.S., and the prescriptions written by them in 2014. This is a subset of the dataset maintained by the Centers for Medicare and Medicaid Services \cite{cms}, that contains almost 24 million Opioid related prescriptions, written by 1 million unique health professionals (prescribers), in the U.S in 2014. Each record in $DS_1$ includes \emph{National Provider Identifier number, provider state, gender, credentials and the number of Opioid related drugs prescribed (among the set of 250 different drugs) by the provider}. In addition, it provides the information whether or not the provider prescribed more or less than 10 Opioid related prescriptions in 2014. It may be noted that determination of whether or not a prescriber has prescribed more than 10 prescriptions in 2014, is not done by summing up the number of drugs prescribed by the provider, as multiple drugs may be prescribed on a single prescription.\\

\textbf{$DS_2$}: This dataset is also collected from \cite{kaggle}. It contains the population in each of the 50 states and also Opioid related deaths in that state. 
\\

\textbf{$DS_3$}: It is the Cincinnati Heroin Overdose dataset available on \cite{cincinnati}. This dataset is a subset of the Emergency Medical Services (EMS) dataset, where each record contains detailed information regarding an incident, such as location, time, EMS response type, neighborhood, and others, that required an EMS dispatch. This dataset contains information related to Heroin incidences from July 2015 to present time. As of April 18, 2018, there were 5568 such incidences. $DS_3$ is a subset of EMS dataset in the sense that it contains information only regarding Heroin incidences. It may be noted that heroin and opioid painkillers are extremely similar in terms of their chemical structure, mechanism of action and range of effects. Accordingly, for the purpose of this study, we use Heroin and other Opioid drug related data, in a similar fashion. \\

\textbf{$DS_4$}: This dataset contains information regarding the median income, median age and educational distribution of various neighborhoods of Cincinnati. Information about the median income, median age and educational distribution were mined from three separate websites \cite{neighborhood,neighborhood2,neighborhood3}.

% We used this dataset, for our classification task. We did not encode the credentials and degree of a prescriber into our model, as wanted to predict the number of prescriptions a \emph{prescriber} would prescribe. \cite{kaggle} contains another dataset, $DS_2$, pertaining to the states in the US, the population of the states and number of Opioid related deaths in that state.\\ 

\subsection{Data Processing and Munging}
\label{munging}
We process and munge data from our collected datasets $DS_1$ through $DS_4$, to create ``secondary'' datasets $DS_5$, $DS_6$ and $DS_7$ to provide answers to the questions raised in Section. \ref{intro}. In the following, we describe these three datasets:\\ \\
\textbf{$DS_5$}: This dataset is created by processing information available in $DS_1$ and $DS_2$. From $DS_1$, we create a temporary dataset $DS_{1.A}$ that contains information regarding the total number of prescribers and prescriptions written in each of the 50 states. $DS_{1.A}$ was \emph{joined} with $DS_2$, to create $DS_5$, that contains information regarding the total number of prescribers, prescriptions and Opioid related deaths in each of the 50 states. \\

\textbf{$DS_6$}: This dataset was created by processing information available in $DS_3$, and it contains information related to the number of Opioid related incidences in each of the 50 neighborhoods of Cincinnati.\\

\textbf{$DS_7$}: This dataset was created by processing information available in datasets $DS_4$ and $DS_6$ and it contains information related to the median income, median age, \emph{median education} and the number of Opioid related incidences in each of the 50 neighborhoods of Cincinnati. It may be noted that $DS_4$ provides information related to the distribution of educational level of each of the neighborhoods. We define median education level of a neighborhood as the number of years, $50\%$ of the residents of the neighborhood spend in school. In \cite{neighborhood2}, the educational level is divided into 10 different categories from $c_1$, ...., $c_{10}$ where $c_1$ corresponds to \emph{None} and $c_{10}$ corresponds to \emph{Doctorate}. The categories $c_1$, ...., $c_{10}$ correspond to $n_1$, ...., $n_{10}$ years of education, with \emph{None} implying 0 years of education and \emph{Doctorate} implying 22 years of education. The precise definition of median education level of a neighborhood is as follows. The median educational level of a neighborhood is $n_k$ years, if $k$ is the smallest integer, such that $\sum_{i = 1}^{k} x_i \geq 50$, where $x_1$, ...., $x_{10}$ represents the percentage of neighborhood population that has educational levels corresponding to $c_1$, ...., $c_{10}$.  \\

\section{Data Analysis Results}
\label{approach}
In this section, we provide results of our data analysis to provide answers to the five questions raised in Section. \ref{intro}. In the following, we discuss the results in detail. \\
\subsection{Data Analysis for Q1}

In order to provide an answer to this question, we first compute the partial correlation \cite{partial} between the number of prescribers in the states with the number of opioid deaths, by controlling the effect of the total number of prescriptions. Next, we compute the partial correlation between the number of prescriptions in the states with the number of opioid deaths, by controlling the effect of the total number of prescribers. Both the correlations were computed using the data available in $DS_5$. It may be noted that partial correlation is a measure of linear relationship between two variables while controlling the effect of a third variable. In this context, we first found the relationship between the prescriber and death by controlling the number of prescriptions and then found the relationship between the number of prescriptions and death by controlling the prescribers. The results of the correlations are presented in the Table. \ref{Partial}. \\
\vspace{-3.00mm}
\begin{table}[h]
\centering
\begin{tabular}{ |c|c|c| }
 \hline
  & Number of  & Number of \\
  & Prescribers & Prescriptions \\ \hline
 Opiate Deaths &  &  \\ 
 (Partial Correlation) & 0.4664 & 0.3619 \\
 \hline
\end{tabular}
\caption{Partial Correlation Coefficients between Opiate Deaths and Prescribers and Prescriptions}
\label{Partial}
\end{table}

In Table. \ref{Partial}, we observe a moderate positive correlation between the number of prescribers and prescriptions with Opioid deaths. This implies that, with an increase in prescribers and prescriptions, there \emph{tends} to be an increase in Opioid related deaths.

\subsection{Data Analysis for Q2}
\begin{figure}[t]
	\begin{center}
		\includegraphics[width=0.50\textwidth, keepaspectratio]{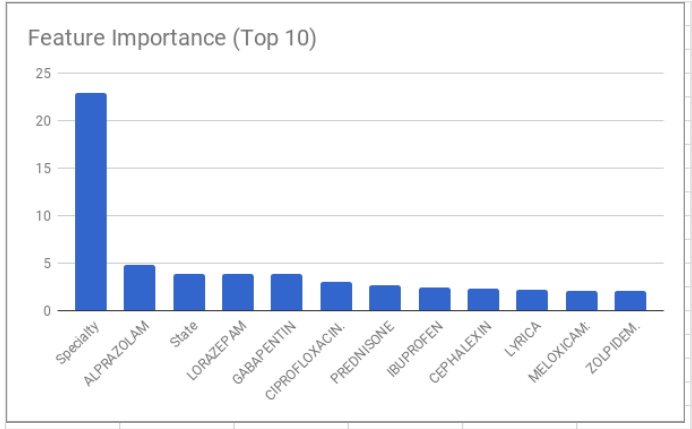}
        \caption{Top 10 Important Features}
		{\label{feature}}
        
	\end{center}
\end{figure}

In the previous section, we analyze the relationship between the prescribers and Opioid-related deaths, and notice that Opioid prescribers play a significant role. Given this information, we try to predict whether a prescriber predicts less than or more than 10 opiate prescriptions in a calendar year, based on $DS_1$ data. This problem can be framed as a supervised classification task with two classes (class 1 and class 2, representing less and more than 10 prescriptions per calender year respectively).\\
As an initial data pre-processing step, we ran a series of boosting algorithms, using data on the non-opioid drugs and treating Gender, State and prescribers’ Specialty as categorical variables. This is because, our motivation was to predict, if a prescriber would prescribe less than or more than 10 Opioid prescriptions, by just analyzing their trend of issuing non Opioid prescriptions.\\
XGBoost \cite{xg} gave a test accuracy of 81.8$\%$. Using CatBoost \cite{cat}, the accuracy increased to 84.7$\%$. The CatBoost algorithm also provided a  feature importance array. A partial list is depicted in Figure \ref{feature}. Providers’ specialty is having the most impact by far. As a follow up, we looked into providers that prescribed Opioids, and their specialties.

We also implemented a multilayer perceptron, for this classification task. The perceptron model considered the State categorical variable along with the non Opioid drugs. The MLP had three hidden layers of sizes, 500, 400 and 300 neurons each. We used the Adam Optimizer to optimize the loss function. The learning rate was set at 0.0001. This model produced a training accuracy of $95.6\%$ and a testing accuracy of $89.7\%$. Finally, we implemented a random forest classifier, with 200 trees. This model had an accuracy of $89\%$. Due to the lack of available data, we could not perform a trend analysis and quantify the growth of the Opioid related deaths, with the prescribers' specialty.

\begin{figure}[t]
	\begin{center}
		\includegraphics[width=0.50\textwidth, keepaspectratio]{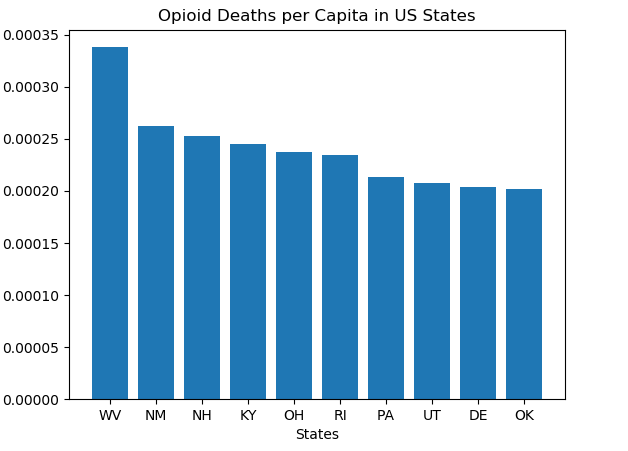}
        \caption{Top 10 Opioid Deaths in US States per Capita}
		{\label{deaths}}
        
	\end{center}
\end{figure}

\begin{figure}[t]
	\begin{center}
		\includegraphics[width=0.50\textwidth, keepaspectratio]{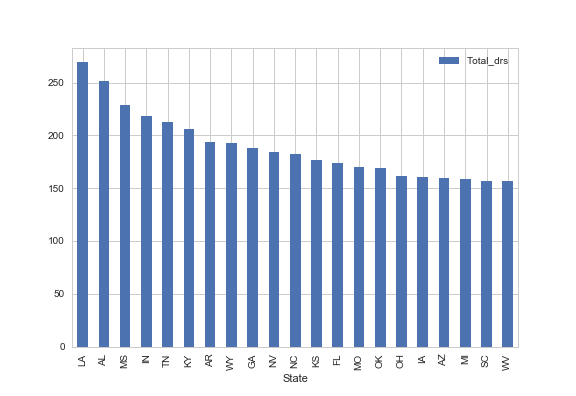}
        \caption{Top 20 High Prescribing States}
		{\label{states}}
        
	\end{center}
\end{figure}

\begin{figure}[t]
	\begin{center}
		\includegraphics[width=0.50\textwidth, keepaspectratio]{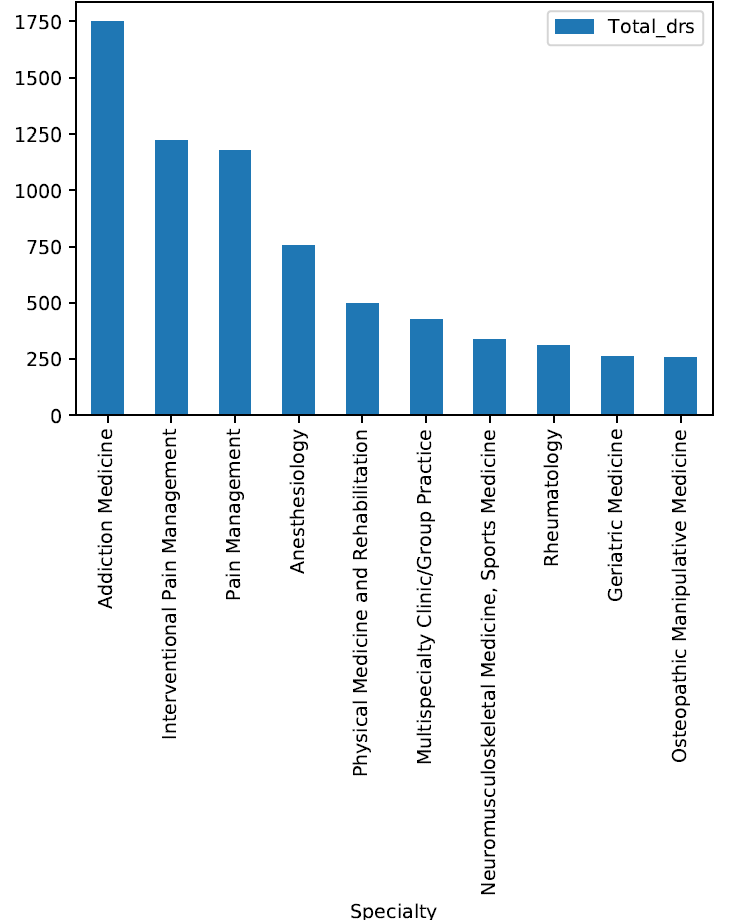}
        \caption{Top 10 High Prescribing Specialty}
		{\label{meds}}
        
	\end{center}
\end{figure}

Having analyzed the role of prescribers in the country, we examined the role of prescribers by state. Adjusting the Opioid related deaths in a state with the population of that state, we get the Opioid deaths per capita, which is shown in Fig. \ref{deaths}. We can see that the state of West Virginia is the worst affected state by this ongoing Opioid epidemic. Fig. \ref{states} illustrates the annual average Opioids prescribed by state, for which there are prescribers who prescribed more than 10 or more Opioids in a year. The figure plots the annual average values for those states which exceed 150 Opioid prescriptions. We next examined the specialties of the prescribers in the country. We discovered that the specialty, Addictive Medicine, prescribed the most average annual Opioid drugs. This result is illustrated in Fig. \ref{meds}. The figure plots the average annual Opioid prescription rate, by specialty, where the number of such prescriptions exceeded 250. Addictive Medicine prescribers usually tend to patients, who are addicted to alcohol, drugs, etc. These results give us an idea of the role of prescribers and their specialties, nationwide and by state, in the ongoing Opioid epidemic. The results are nothing but a starting point in the mitigation of the epidemic, using data analytic techniques.    

\subsection{Data Analysis for Q3}

In order to provide an answer to this question, we first compute the partial correlation \cite{partial} between the median income in a neighborhood to the number of Opioid related deaths in that neighborhood, by controlling the effect of median age and median education level of the neighborhood. The correlation was computed using the data available in $DS_7$.  The result of this correlation can be found in column 2 of Table. \ref{PartialCin}. \\
This result shows us that the median income of the neighborhoods have a very small negative correlation with the Opioid related deaths in the neighborhoods. This implies that, the Opioid addiction has spread to all income levels, in the neighborhoods of Cincinnati.

\begin{table}[h!]
\centering
\begin{tabular}{ |c|c|c|c|}
 \hline
  & Median & Median & Median   \\ 
  &  Income & Age & Education \\\hline
 Opiate Deaths &  & &\\ 
 (Partial Correlation) & -0.0576& -0.0789 & -0.1516  \\
 \hline
\end{tabular}
\caption{Partial Correlation Coefficients between Opiate Deaths and Median Income, Median Age and Median Education}
\label{PartialCin}

\end{table}

\subsection{Data Analysis for Q4}

In order to provide an answer to this question, we first compute the partial correlation between the median age in a neighborhood to the number of Opioid related deaths in that neighborhood, by controlling the effect of median income and median education level of the neighborhood. The correlation was computed using the data available in $DS_7$.  The result is presented in column 3 of Table. \ref{PartialCin}. \\
The partial correlation coefficient of median age and Opioid related deaths is similar to that of median income and Opioid related deaths. This is small negative correlation illustrates that the addiction is not just concentrated to certain age groups, but affects individuals of all ages.

\subsection{Data Analysis for Q5}
In order to provide an answer to this question, we first compute the partial correlation between the median education in a neighborhood to the number of Opioid related deaths in that neighborhood, by controlling the effect of median income and median age level of the neighborhood. The correlation was computed using the data available in $DS_7$.  The result is presented in column 4 of Table. \ref{PartialCin}. \\
This result illustrates that there is a weak negative partial correlation between the median education and Opioid related deaths. This implies that, as the median education value increases, the Opioid related deaths \emph{tend} to decrease. 

\subsection{Joint Analysis for Q3, Q4 and Q5}
In order to determine the joint effect of the median income, median age and median education level of the neighborhoods on Opioid related deaths, we computed the multiple correlation between these three predictor variables and the target variable, Opioid related deaths. We found a weak positive correlation, which implies that median income, median age and median education are related in a way that is producing a counter-intuitive result. To decipher the accurate impact of these three predictor variables on the target variable, we require more granular level data. This granular data will help to model these variables and to calculate the impact of these variables on Opioid-related deaths.

\section{Conclusions}
\label{Conclusions}
Firstly, we highlight the role of prescribers and prescription opiate drugs, by analyzing their role with the number of Opioid related deaths in 2014. This analysis shows that there is a moderate positive correlation between the number of prescribers and the number of prescriptions with the number of Opioid related deaths in U.S. states. Secondly, our classification models report higher accuracy when compared to the benchmark scores of IBM. We analyzed the possibility of a prescriber prescribing Opiate drugs, by studying their trend of issuing non Opioid prescriptions. Thirdly, we take a look at the neighborhoods of Cincinnati to observe the impact of income, age and education on the Opioid related deaths in the city. We find that the Opioid addiction affects individuals of all income and age levels, and is not just limited to one specific level. Finally, we observe that, with an increase in the educational levels of a neighborhood, the Opioid related deaths \emph{tend} to decrease.

\end{document}